\documentclass[aps,prd,twocolumn]{revtex4-1}
\usepackage{hyperref}
\usepackage{graphicx}
\usepackage{amsmath}
\usepackage[utf8]{inputenc}

\begin{document}

\title{Can a pure vector gravitational wave mimic a pure tensor one?}

\date{\today}

\author{Bruce Allen}
\affiliation{MPI for Gravitational Physics, Callinstrasse 38, Hannover, Germany}

\begin{abstract}
  \noindent
  In the general theory of relativity, gravitational waves have two
  possible polarizations, which are transverse and traceless with
  helicity $\pm 2$.  Some alternative theories contain additional
  helicity $0$ and helicity $\pm 1$ polarization modes.  Here, we
  consider a hypothetical ``pure vector'' theory in which
  gravitational waves have only two possible polarizations, with
  helicity $\pm 1$.  We show that if these polarizations are allowed
  to rotate as the wave propagates, then for certain source locations
  on the sky, the strain outputs of three ideal interferometric
  gravitational wave detectors can exactly reproduce the strain
  outputs predicted by general relativity.
\end{abstract}

DOI: 10.1103/PhysRevD.97.124020


\maketitle

\section{Introduction}
\label{s:intro}

In Einstein's general theory of relativity (GR), gravitational waves
(GW) are radiative perturbations in the geometry of space-time that
propagate at the speed of light \cite{Einstein:1916cc,
  Einstein:1918btx}. These so-called ``tensor'' modes are spin-2
transverse traceless metric perturbations, which have only helicity
$\pm 2$ degrees of freedom \cite{Wald:1984rg}.  In alternatives to GR,
GWs can contain additional helicity $\pm 1$ and helicity 0 degrees of
freedom, which are often called ``vector'' and ``scalar'' modes
\cite{Misner:1974qy, Schutz:1985jx, Will:2014kxa}.

Interferometric GW detectors such as the Laser Interferometer
Gravitational-wave Observatory (LIGO) \cite{TheLIGOScientific:2014jea}
and Virgo \cite{TheVirgo:2014hva} would respond differently to vector
and scalar modes than they respond to tensor modes.  When more than
two GW detectors~\footnote{We assume that these detectors can only
  observe a single polarization mode, or equivalently only a single
  linear combination of polarization modes.} observe the same source,
this makes it possible, at least in principle, to measure the wave
polarization content and determine which helicity components are
present \cite{Cadonati:2004ms, Wen:2005ui, Isi:2017equ,
  Eardley:1973br, Eardley:1974nw, Isi:2015cva, Nishizawa:2009bf}.

On 14 August 2017, the two LIGO detectors in the USA and the Virgo
detector in Italy made the first simultaneous 3-detector direct
observation of GWs \cite{Abbott:2017oio}; the signal was present in
the band of the detectors for a fraction of a second.  The waves were
emitted by the merging binary black hole system GW170814, and excited
the three detectors in a pattern that was consistent with the pure
tensor mode predictions of GR.  The discovery paper
\cite{Abbott:2017oio} also claimed that the observation strongly
disfavored a pure vector model, which would not match the
excitation pattern observed in the detectors.

The presentation in \cite{Abbott:2017oio} suggests that this is a
theory-independent observational result.  Here we show that this is
not the case, by explicitly constructing a pure vector
gravitational wave which for certain source locations on the sky can
precisely mimic or match the predictions of GR.  We assume that the
signal has short duration compared to the 24h rotation period of the
Earth, so that during the observation the detector orientations
relative to the line of sight to the source do not change
significantly.

To construct this counterexample, we allow the ``direction'' of the
wave polarization to change as the GW propagates, in the same way that
the Faraday effect can rotate the polarization of an electromagnetic
wave~\cite{Jackson:1998nia, Fowles:1968zgx, 1979rpa..book.....R,
  1999prop.book.....B}.  With an appropriate amount of rotation, the
GW remains pure vector but exactly mimics the detector response
expected for tensor modes in GR. 

The paper is organized as follows.  In Sec.~\ref{s:planewavesinGR}
we discuss the ``pure tensor'' weak-field plane wave solution of GR.
In Sec.~\ref{s:3detectors} we show how an array of three
interferometric detectors would respond to these waves.  In
Sec.~\ref{s:3detinPVG} we then consider the same questions for a
conventional pure vector theory, and discuss how null
streams~\cite{Wen:2005ui} can distinguish these from conventional GR.
In Sec.~\ref{s:rotatingbasis} we construct an orthonormal basis for
polarization tensors which rotates about the propagation axis. In
Sec.~\ref{s:counterexample} we use the rotating basis to construct
a pure vector gravitational wave which changes its polarization as
it propagates (behavior that does not happen in GR).  By correctly
tuning the amount of rotation, we show that for some sky positions,
this pure vector theory would produce exactly the same detector
outputs as GR.  This is followed by a short conclusion, and an
Appendix giving quantities needed to reproduce the results, such as
the antenna response tensors for the LIGO and Virgo detectors.

\section{Plane waves in general relativity}
\label{s:planewavesinGR}
In GR, a GW propagating in the positive $z$-direction may be written
in the weak field limit as
\begin{equation}
  h_{ab} = w_+(t - z/c) e^+_{ab} + w_\times(t - z/c) e^\times_{ab}.
  \label{e:planewave}
\end{equation}
Here $c$ is the speed of light and $w_+$ and $w_\times$ are arbitrary
real functions of one variable, which are the waveforms of the two
different polarization modes.  The function $t-z/c$ is called
``retarded time''~\footnote{This retarded time is not referred to the
  source, but rather to the time at which the corresponding wavefront
  passes the plane defined by $z=0$.}.  The transverse traceless
polarization tensors are
\begin{equation}
  \begin{aligned}
  e^+_{ab}     & =  x_a x_b - y_a y_b \\
  e^\times_{ab} & =  x_a y_b + y_a x_b
  \end{aligned}
  \label{e:spin2polbasisvectors}
\end{equation}
where $x^a$ and $y^a$ are unit vectors~\footnote{Since we are in the
  weak field limit, indices may be freely raised and lowered.} in the
x and y directions.  The complex combinations $e^+_{ab} \pm i
e^\times_{ab}$ have helicity $\pm 2$, because under rotation of the
orthogonal pair $x^a, y^a$ through angle $\phi$ in the plane orthogonal
to the propagation direction $z$, the complex polarizations acquire
phase factors $\exp(\pm 2 i \phi)$ \cite{Feynman:1996kb}.

The strain response $h(t)$ of an ideal interferometric gravitational
wave detector, with perpendicular arms of equal length is
\begin{equation}
  h(t) = d^{ab} h_{ab}(t, z_0),
  \label{e:howdetresponds}
\end{equation}
where $z_0$ is the z coordinate of the detector's location, and
\begin{equation}
  d^{ab} = L_1^a L_1^b - L_2^a L_2^b
  \label{e:detresponsetensor}
\end{equation}
is the detector response tensor.  The quantities $ L_1^a$ and $ L_2^a$
are unit vectors along the two detector arms, and the detector is
assumed to be ideal. This means that it is free of instrumental noise,
with a flat frequency response that is broad enough to reproduce the
spectral content of the waveforms $w^+$ and $w^\times$.

\section{The case of three detectors}
\label{s:3detectors}

Now suppose that we have three ideal detectors, at locations with z
coordinate $z_i$ and with detector response tensors $d^{ab}_i$, for
$i=1, \cdots, 3$.  We assume that the GW source is at a distance from
the detectors which is much larger than their size, than the
separation between them, and than the characteristic wavelength of the
radiation.  We assume that the source's sky location is known
precisely, and (without loss of generality) that it lies on the
negative $z$ axis.  We also assume that the signal is visible in the
detectors for a time that is brief compared to the Earth's spin period
of 24h, so that the instrument orientations may be treated as fixed.

In this case, the strain at the three detectors is given by combining
(\ref{e:planewave}) and (\ref{e:howdetresponds}):
\begin{equation}
  h_i(t) = F^+_i w_+(t - z_i/c)+ F^\times_i w_\times(t - z_i/c) .
\end{equation}
The strain is seen to be a linear combination of the two different
waveforms, with weights $F_i^+$ and $F_i^\times$ that are determined
by the geometrical overlap between the detector response tensors and
the polarization tensors. These weights,
\begin{equation}
  F_i^+ = e^+_{ab} d_i^{ab} \quad \text{and} \quad F_i^\times =
  e^\times_{ab} d_i^{ab},
\end{equation}
are called ``antenna functions'' and are easily computed.  Here, they
should be thought of as dimensionless numbers, whose fixed values are
determined \emph{a priori} by the orientations of the detector arms
\cite{GR-QC/9607075} relative to the GW source.  (For completeness,
the Appendix gives expressions for $d_i^{ab}$ for the
LIGO and Virgo detectors.)

For later convenience, we introduce the non-negative lengths $F_i$
of these three two-vectors
\begin{equation}
    F_i = \sqrt{(F^+_i)^2 + (F^\times_i)^2}
\end{equation}
and the angles defined by the ratios of the two components,
\begin{equation}
  \phi_i = \arctan \frac{F^\times_i}{F^+_i}.
\end{equation}
Here, the arc-tangent is defined as the principal value of $\arg(F^+ +
iF^\times)$.  In terms of these lengths and angles, we can write the
two-vectors $(F^+_i, F^\times_i) = F_i (\cos \phi_i, \sin \phi_i)$.
As before, $F_i$ and $\phi_i$ should be thought of as constants, which
are defined by the detector arm orientations relative to the direction
to the source.

To compare the strain outputs of the different detectors, it is
helpful to first shift the observed strains to a common fiducial
arrival time, defining new time series by $\bar h_i(t) = h_i(t +
z_i/c)$. In GR these time-shifted strains may be written as linear
combinations of the two waveforms, with weights determined by the
antenna patterns of the three instruments,
\begin{equation}
  \left[
    \begin{array}{c}
      \bar h_1(t) \\
      \bar h_2(t) \\
      \bar h_3(t)
    \end{array}
    \right] =
  \begin{array}{c}
    F_1 \\
    F_2 \\
    F_3
  \end{array}
  \left[
    \begin{array}{cc}
      \cos \phi_1  & \sin \phi_1 \\
      \cos \phi_2  & \sin \phi_2 \\
      \cos \phi_3  & \sin \phi_3
    \end{array}
    \right]
  \left[
    \begin{array}{c}
      w_+(t) \\
      w_\times(t)  \\
    \end{array}
    \right],
  \label{e:3x2matrixinGR}
\end{equation}    
where the respective values of $F_i$ are understood to multiply the
corresponding row on the right-hand side (r.h.s.).

For generic detector orientations, the three two-vectors $(F^+_i,
F^\times_i)$ are nonvanishing.  Hence they must be linearly
dependent, since a two-dimensional vector space can contain at most
two linearly independent vectors.  Thus, the three time-shifted
strains must also be linearly dependent, with time-independent
coefficients, since they are linear combinations of two independent
waveform functions $w_+(t)$ and $w_\times(t)$.

This is the basis of ``null tests'', which construct a linear
combination of these three time-shifted strains $\bar h_1(t), \bar
h_2(t), \bar h_3(t)$ which vanishes if GR is a correct description of
GW \cite{Wen:2005ui}.  This vanishing linear combination is called a
``null stream''.  By the same reasoning, if there are $N$
generically-oriented detectors, then $N-2$ ``different'' null stream
combinations may be constructed from them~\cite{Wen:2005ui}.

\section{The three-detector case in ``pure vector gravity''}
\label{s:3detinPVG}
Now consider a fictitious pure vector theory of gravity, where
the transverse traceless polarization tensors of GR are replaced by a
pair which for convenience we denote with $L$ and $R$:
\begin{equation}
  \begin{aligned}
  e^L_{ab} & = x_a z_b + z_a x_b \\
  e^R_{ab} & = y_a z_b + z_a y_b
  \label{e:LRbasis}
  \end{aligned}
\end{equation}
where $z^a$ is a unit vector in the positive $z$ direction.  In this
pure vector case, the GW is described by
\begin{equation}
  h_{ab} = w_L(t - z/c) e^L_{ab} + w_R(t - z/c) e^R_{ab},
  \label{e:pvwave}
\end{equation}
where, as before, the waveforms $w_L$ and $w_R$ are arbitrary real
functions of one variable.

We assume that in this fictitious theory of gravity, the detectors
respond as before (\ref{e:howdetresponds}) so that the strain in the
$i$th detector is
\begin{equation}
  h_i(t) = F^L_i w_L (t - z_i/c)+  F^R_i w_R(t - z_i/c),
\end{equation}
where the weights $F$ that appear are again the overlaps between
the detector response tensors and the polarization tensors:
\begin{equation}
  F_i^L = e^L_{ab} d_i^{ab} \quad \text{and} \quad F_i^R = e^R_{ab}
  d_i^{ab}.
  \label{e:vectorpols}
\end{equation}
As before, after shifting to a common fiducial time, the three time
series $\bar h_i(t) = h_i(t+z_i/c)$ must be linearly dependent,
because they are linear combinations of the two functions $w_L(t)$ and
$w_R(t)$.

As in the previous section, we introduce the non-negative lengths
$f_i$ of these three two-vectors
\begin{equation}
    f_i = \sqrt{(F^L_i)^2 + (F^R_i)^2}
\end{equation}
and the angles defined by the ratios of the two components,
\begin{equation}
  \psi_i = \arctan \frac{F^R_i}{F^L_i},
\end{equation}
where the arc-tangent is defined as the principal value of $\arg(F^L +
iF^R)$.  In terms of these quantities, we can express the two-vectors
as $(F^L_i, F^R_i) = f_i (\cos \psi_i, \sin \psi_i)$, where again
$f_i$ and $\psi_i$ are constants defined by the detector arm
orientations relative to the direction to the source.

The time-shifted strains resulting from this pure vector theory are then
\begin{equation}
  \left[
    \begin{array}{c}
      \bar h_1(t) \\
      \bar h_2(t) \\
      \bar h_3(t)
    \end{array}
    \right] =
  \begin{array}{c}
    f_1 \\
    f_2 \\
    f_3
  \end{array}
  \left[
    \begin{array}{cc}
      \cos \psi_1  & \sin \psi_1 \\
      \cos \psi_2  & \sin \psi_2 \\
      \cos \psi_3  & \sin \psi_3
    \end{array}
    \right]
  \left[
    \begin{array}{c}
      w_L(t) \\
      w_R(t)  \\
    \end{array}
    \right],
  \label{e:PVmatrix0}
\end{equation}    
where the respective values of $f_i$ are understood to multiply the
$i$th row on the r.h.s.  This should be compared with the
corresponding Eq.~(\ref{e:3x2matrixinGR}) that arises in GR.

For generic detector orientations, the first two rows of this system
are invertible, so the data at the first two detectors completely
determines $w_L(t)$ and $w_R(t)$. With that choice, $\bar h_1(t)$ and
$\bar h_2(t)$ will match the data perfectly.  But if GR is correct,
then $\bar h_3(t)$ in (\ref{e:PVmatrix0}) will be incorrect: it will
not match the data from the detector.

\section{A rotating basis for polarization tensors}
\label{s:rotatingbasis}

At this juncture, it is helpful to introduce a ``rotating'' vector
polarization basis~\footnote{This basis rotates in phase at frequency
  ``one cycle per rotation'' ($\phi$) as opposed to (say) frequency
  ``two cycles per rotation'' ($2\phi$) or ``no cycles per rotation''
  ($0\phi$).  Since the polarization basis transforms as a rank-(0,2)
  tensor, this implies that the vector polarizations are helicity $\pm
  1$ components of a spin-2 field.  The name ``vector'' perturbations
  is therefore somewhat misleading, since this implies a spin-1
  object, which is a rank-(0,1) or (1,0) tensor.  However it has
  historically been used in this context, so we perpetuate the
  misnomer.}  analogous to (\ref{e:LRbasis}).  For this, we first
define a basis that rotates with space and time as
\begin{equation}
  \begin{aligned}
  x'^a   & =  \cos \phi(t,z) x^a  + \sin \phi(t,z) y^a \\
  y'^a   & =  - \sin \phi(t,z) x^a + \cos \phi(t,z) y^a \\
  z'^a   & =  z^a \\
  \end{aligned}
  \label{e:phasedef}
\end{equation}
where the phase $\phi$ is an arbitrary real function of time $t$ and
spatial position $z$.  The new basis vectors $(x'^a, y'^a)$ are
obtained by rotating the old basis vectors $(x^a,y^a)$
counterclockwise through angle $\phi$.

Using this basis, we define new polarization tensors by
\begin{equation}
  \begin{aligned}
  e^l_{ab} & =  x'_a z'_b + z'_a x'_b = \cos \phi \, e^L_{ab} + \sin \phi  \, e^R_{ab} \\
  e^r_{ab} & =  y'_a z'_b + z'_a y'_b  = - \sin \phi \, e^L_{ab} +
  \cos \phi  \, e^R_{ab}
  \end{aligned}
  \label{e:pol}
\end{equation}
and express the gravitational wave amplitude as a sum over these two
polarizations:
\begin{equation}
  \begin{aligned}
    h_{ab}(t,z) & = w_l(t - z/c) e^l_{ab} + w_r(t - z/c) e^r_{ab}.
  \label{e:hab}
  \\
  & \\
  \end{aligned}
\end{equation}
As before, the waveform functions $w_l$ and $w_r$ are arbitrary real
functions of one variable.  Note however that the left-hand side is a function of
retarded time $t-z/c$ if and only if the phase $\phi$ is {\it also} a
function of retarded time.

We stress that (\ref{e:phasedef}) is \emph{not} defining a new basis
for the tensor space over our space-time manifold.  If it were, then
scalar quantities such as the strain $h$ at a particular detector
would not be changed.  This is because the components of the field
$h_{ab}$ would transform covariantly, but the components of the
detector tensors $d^{ab}$ would transform contravariantly, and the
scalar product $h_{ab}d^{ab}$ would be invariant.  Here, we are
explicitly \emph{changing} the nature and physical content of the GW
in ways that we will discuss shortly.

By using (\ref{e:pvwave}), (\ref{e:vectorpols}), (\ref{e:pol}), and
(\ref{e:hab}), we can express the time-shifted detector strains as the
2-dimensional vector inner products
\begin{widetext}
\begin{equation}
  {\bar h}_i(t) = \left[ F^L_i \quad F^R_i \right]
  \left[
  \begin{array}{cc}
    \cos \phi(t+z_i/c, z_i)  & -\sin \phi(t + z_i/c, z_i) \\
     \sin \phi(t+z_i/c, z_i) & \cos \phi(t + z_i/c, z_i)
  \end{array}
  \right]
  \left[
  \begin{array}{c}
  w_l(t) \\
  w_r(t)
  \end{array}
  \right].
  \label{e:changingpolarizations}
\end{equation}
\end{widetext}
If the phase $\phi$ only depends upon (a single variable, which is)
the retarded time $t-z/c$, then this is completely equivalent to the
case considered in the previous section with waveform functions
$w_L(t)$ and $w_R(t)$ defined by
\begin{equation}
 \left[ \begin{array}{c}
  w_L(t) \\
  w_R(t)
  \end{array} \right] = 
  \left[
  \begin{array}{cc}
    \cos \phi(t) & - \sin \phi(t) \\
    \sin \phi(t) & \cos \phi(t)
  \end{array}
  \right]
  \left[
  \begin{array}{c}
  w_l(t) \\
  w_r(t)
  \end{array}
  \right].
  \label{e:completelyequivalent}
\end{equation}
As before, with three strain functions observed at the detectors, but
only two functional degrees of freedom available, the system is
overconstrained.

\section{Constructing a counterexample}
\label{s:counterexample}

In electrodynamics, when a wave is propagating in vacuum, then the
polarization state is ``carried'' by the wave-fronts.  For an observer
at a fixed point in space, the direction of the electric field
polarization vector may change with time. But if we consider the
polarization vector along the null path followed by the wave, then the
polarization is constant along that path.  In this sense, the
wavefront carries the polarization state with it
\cite{1995ocqo.book.....M, 1999prop.book.....B}. However this is only
true in a vacuum.

If instead, the wave is propagating through a material, then the
polarization vector can shift direction along the path followed by the
wavefront.  One example is Faraday rotation, when the wave is
propagating through a medium containing free charges in the presence
of a magnetic field \cite{Jackson:1998nia, 1979rpa..book.....R,
  Fowles:1968zgx}.

GR is similar to vacuum electrodynamics, where by ``polarization'' we
mean the relative ``mixture'' of $e^+_{ab}$ and $e^\times_{ab}$ that
is present in the wave, for a fixed Cartesian basis.  For a given
fixed observer, this mixture will vary with time as the GW passes
by. But if we look along the null path traced by a point on the
wavefront, then it follows from (\ref{e:planewave}) that the mixture
does not change. However there is no reason to assume that the
conjectured pure vector theory of gravity obeys this way!

To construct a pure vector gravitational wave which gives rise to
the same strains as would be found in pure tensor GR, we assume
that in the pure vector theory (for unspecified reasons) the
polarization state can \emph{change} as the wave propagates.  For
example, this might be due to the presence of the Earth, which
hypothetically could introduce a dependence on the variable $z$, so
that
\begin{equation}
  \phi(t, z) = \Delta \psi(z).
\end{equation}
As can be seen from Eqs.~(\ref{e:PVmatrix0}) and
(\ref{e:completelyequivalent}), the effect is equivalent to rotating
the antenna pattern function through an additional angle $\Delta
\psi_i = \Delta \psi(z_i)$ at the $i$th detector.  Defining $\tilde
\psi_i = \psi_i + \Delta \psi_i$, the response of the detectors will
now be described by
\begin{equation}
  \left[
    \begin{array}{c}
      \bar h_1(t) \\
      \bar h_2(t) \\
      \bar h_3(t)
    \end{array}
    \right] =
  \begin{array}{c}
    f_1 \\
    f_2 \\
    f_3
  \end{array}
  \left[
    \begin{array}{cc}
      \cos \tilde \psi_1  & \sin \tilde\psi_1 \\
      \cos \tilde \psi_2  & \sin \tilde\psi_2 \\
      \cos \tilde \psi_3  & \sin \tilde\psi_3
    \end{array}
    \right]
  \left[
    \begin{array}{c}
      w_L(t) \\
      w_R(t)  \\
    \end{array}
    \right],
\label{e:PVmatrix}
\end{equation}
where the factors $f_i$ multiply the corresponding lines on the r.h.s.

Now that we have introduced additional degrees of freedom in the
$\tilde \psi_i$, it will be possible \emph{in some cases} for this
hypothetical pure vector theory to exactly match the strains
predicted or expected for GR.  However this is not always possible.
Roughly speaking, by changing the polarization angle shifts $\Delta
\psi_i$ , we can vary the proportions of the $w_R$ and $w_L$ waveforms
that enter the detected strain at each site.  However if the vector
response $f_i$ at a particular detector is too small compared to the
GR response at that detector, then it won't be possible to match the
GR prediction, because varying only the phase $\Delta \psi_i$ can not
make the amplitude arbitrarily large.  As an extreme example, suppose
that the source is located at a position in the sky for which the
third detector has no response to {\it either} of the pure vector
modes, which means that $f_3$ vanishes.  In this case there is no way
for the pure vector theory to give the same observed strain as GR
at the third detector~\footnote{We assume here that $F_3$ does not
  vanish, meaning that the GR strain in the third detector is
  nonvanishing.}

Can we pick $\Delta \psi_i$ so that the pure vector theory
masquerades as ordinary GR?  It turns out that this is
possible if and only if the antenna patterns obey a triangle
inequality.  This will only be satisfied for some incoming wave
directions.

To understand what directions, invert the first two rows of
Eq.~(\ref{e:3x2matrixinGR}) to obtain $w_+$ and $w_\times$ as
linear functions of $\bar h_1$ and $\bar h_2$. Then use the first two
rows of the pure vector Eq.~(\ref{e:PVmatrix}) to replace
$\bar h_1$ and $\bar h_2$, obtaining expressions for $w_+$ and
$w_\times$ as linear functions of $w_L$ and $w_R$.  With these
choices, we have ensured that the pure vector model will produce
strains at detectors 1 and 2 which agree with those of GR.

Can we pick the $\Delta \psi_i$ (or equivalently, $\psi_i$) so that
the strain at the third detector would agree for GR and the pure
vector model?  This implies that the third rows of
(\ref{e:3x2matrixinGR}) and (\ref{e:PVmatrix}) agree for all choices
of $w_R$ and $w_L$, which is possible if and only if
\begin{equation}
  f_3 (\cos \tilde \psi_3 w_L + \sin \tilde \psi_3 w_R) = F_3 (\cos \phi_3 w_+ + \sin \phi_3 w_\times).
\end{equation}
Replacing $w_+$ and $w_\times$ on the r.h.s. with the linear
combinations of $w_L$ and $w_R$ obtained above, multiplying by a
factor of $F_1 F_2 \cos(\phi_2-\phi_1)$, and collecting terms, we
obtain
\begin{widetext}
\begin{equation}
  \begin{aligned}
    w_L(t) & \bigl[
      f_1 F_2 F_3 \sin(\phi_3-\phi_2) \cos \tilde \psi_1 +
      F_1 f_2 F_3 \sin(\phi_1-\phi_3) \cos \tilde \psi_2 +
      F_1 F_2 f_3 \sin(\phi_2-\phi_1) \cos \tilde \psi_3 \bigr] + \\
    w_R(t) & \bigl[
      f_1 F_2 F_3 \sin(\phi_3-\phi_2) \sin \tilde \psi_1 +
      F_1 f_2 F_3 \sin(\phi_1-\phi_3) \sin \tilde \psi_2 +
      F_1 F_2 f_3 \sin(\phi_2-\phi_1) \sin \tilde \psi_3 \bigr]=0.
  \end{aligned}      
\end{equation}
\end{widetext}
If it is possible to pick $\Delta \psi_i$ (or equivalently $\tilde
\psi_i$) so that both of the quantities in square brackets vanish,
then the pure vector theory can masquerade as GR.

The problem may be visualized as follows.  Each of the three $\tilde
\psi_i$ define a unit-length vector $\hat n_i = (\cos \tilde \psi_i,
\sin \tilde \psi_i)$ in the $(L,R)$ plane.  The quantities in square
brackets will vanish if and only if the directions of these three unit
vectors can be chosen so that the sum
\begin{equation}
  \begin{aligned}
  & f_1 F_2 F_3 \sin(\phi_3-\phi_2) \hat n_1 + F_1 f_2 F_3
    \sin(\phi_1-\phi_3) \hat n_2 \\ + &  F_1 F_2 f_3
    \sin(\phi_2-\phi_1) \hat n_3
    \end{aligned}
\end{equation}
vanishes.  This in turn is possible if and only if the three lengths
that multiply these unit vectors obey the triangle inequality: each of
the three must not be larger than the sum of the other two.

If we use the definitions of the $\phi_i$, then this can be
re-expressed in terms of the original antenna pattern weights.  This
postulated pure vector theory can masquerade as GR provided that
the three cyclically-related quantities
\begin{equation}
    \begin{aligned}
      & f_1 (F_2^+ F_3^\times - F_2^\times F_3^+), \\
      & f_2 (F_3^+ F_1^\times - F_3^\times F_1^+),  \\
  \text{and} \quad    & f_3 (F_1^+ F_2^\times - F_1^\times F_2^+)
  \end{aligned}
\end{equation}
obey the triangle inequality: the sum of any two must be greater than
or equal to the third.

Note that if a particular incoming wave direction $\hat z$ satisfies
this inequality, then so does the antipodal direction $-\hat z$.  This
is because, under the antipodal transformation $\vec x \rightarrow -
\vec x$, the antenna weights $F^+_i$ and $F^\times_i$ are invariant,
as are the $f_i$.  So a sky map of the regions where the triangle
inequality is satisfied is symmetric under antipodal transformations.

\begin{figure*}
  \begin{center}
  \begin{tabular}{cc}
    \includegraphics[width=0.4\textwidth]{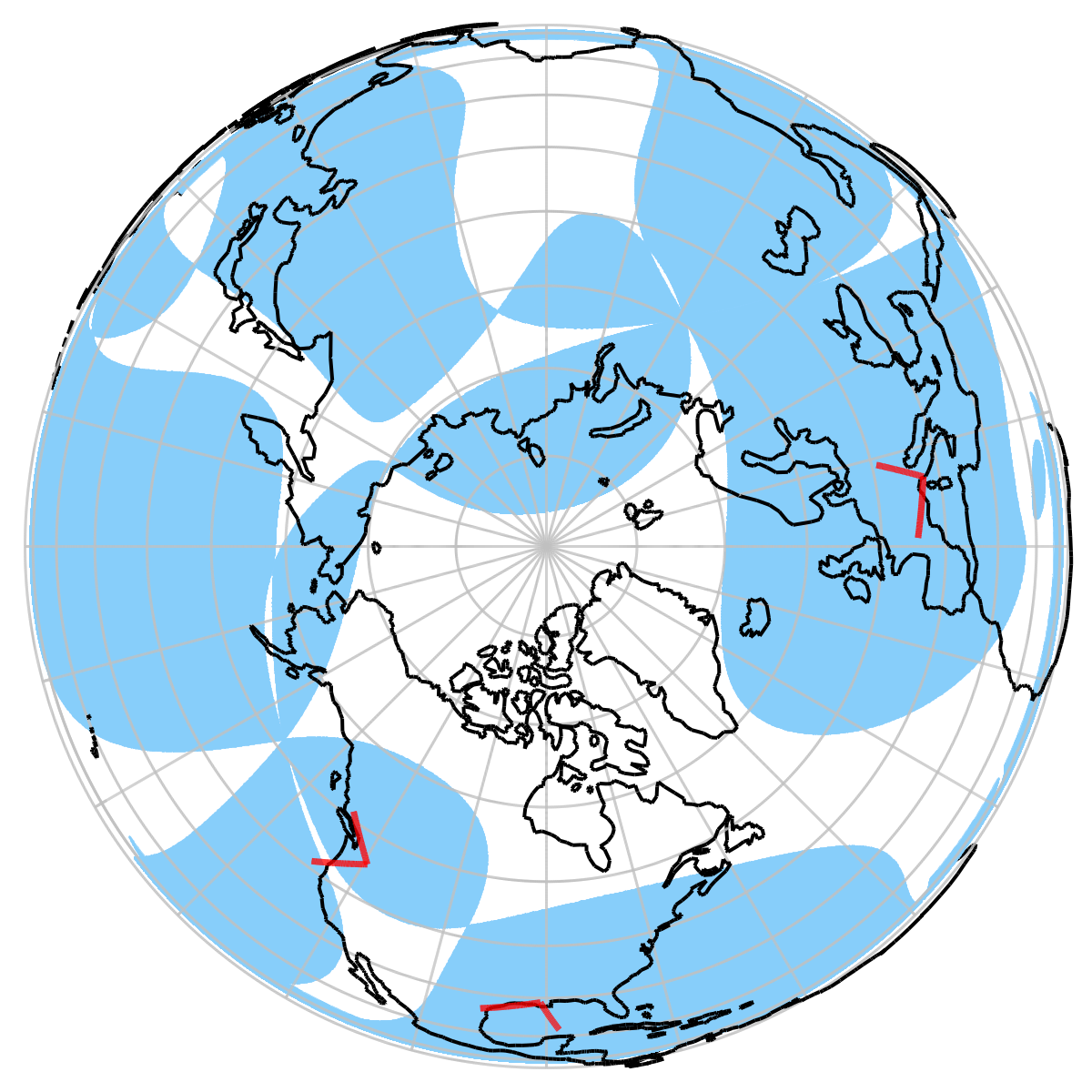} & \includegraphics[width=0.4\textwidth]{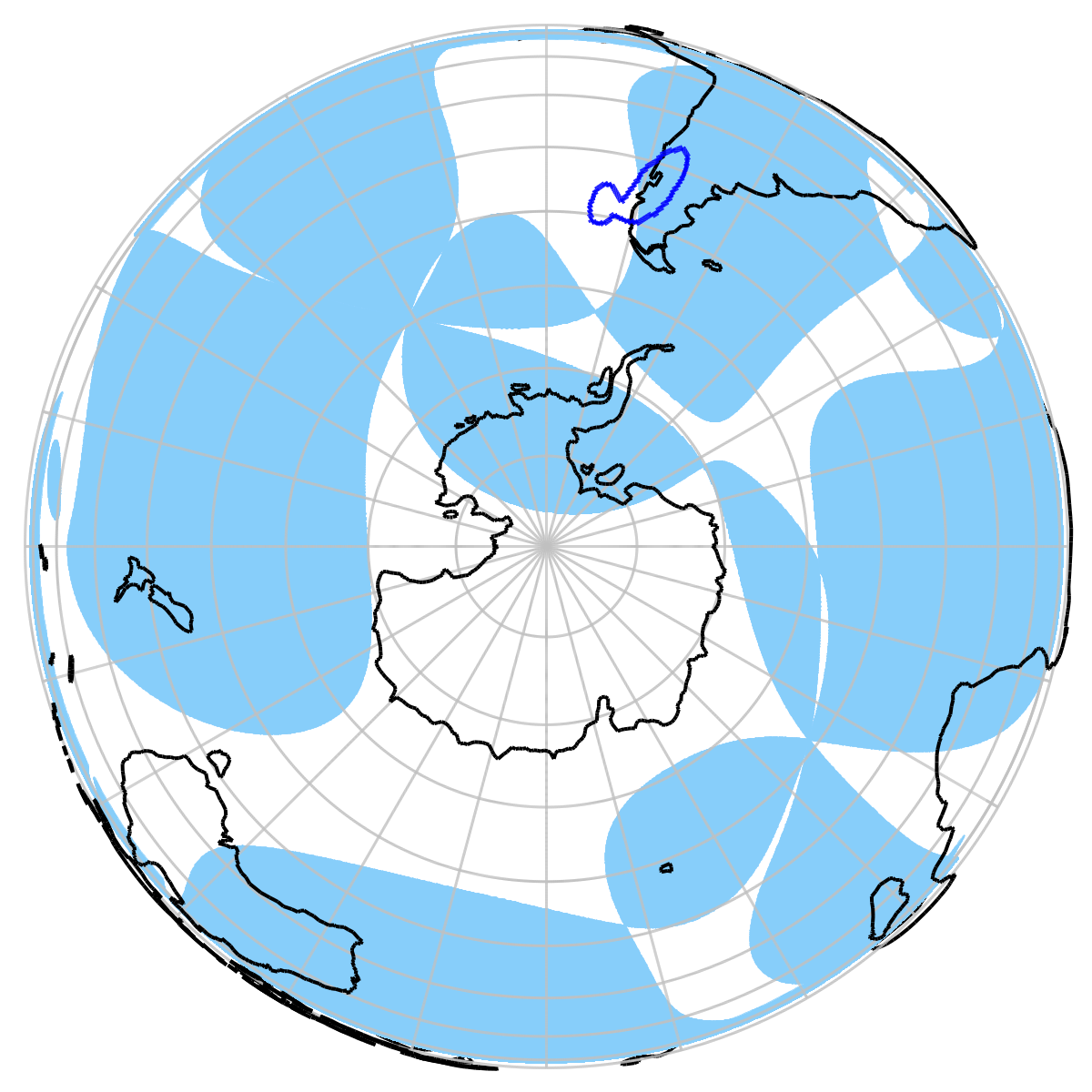} \\
    \includegraphics[width=0.4\textwidth]{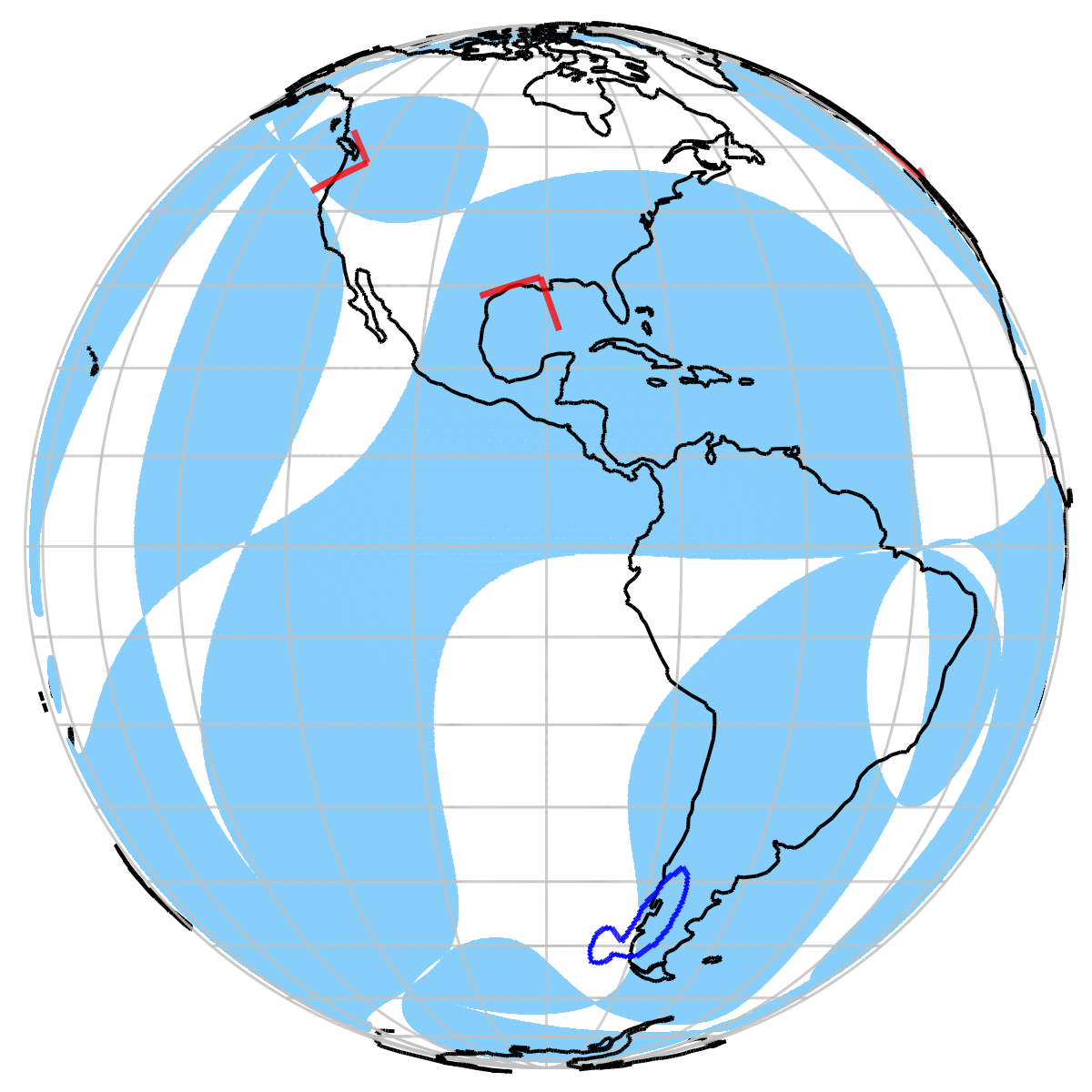} & \includegraphics[width=0.4\textwidth]{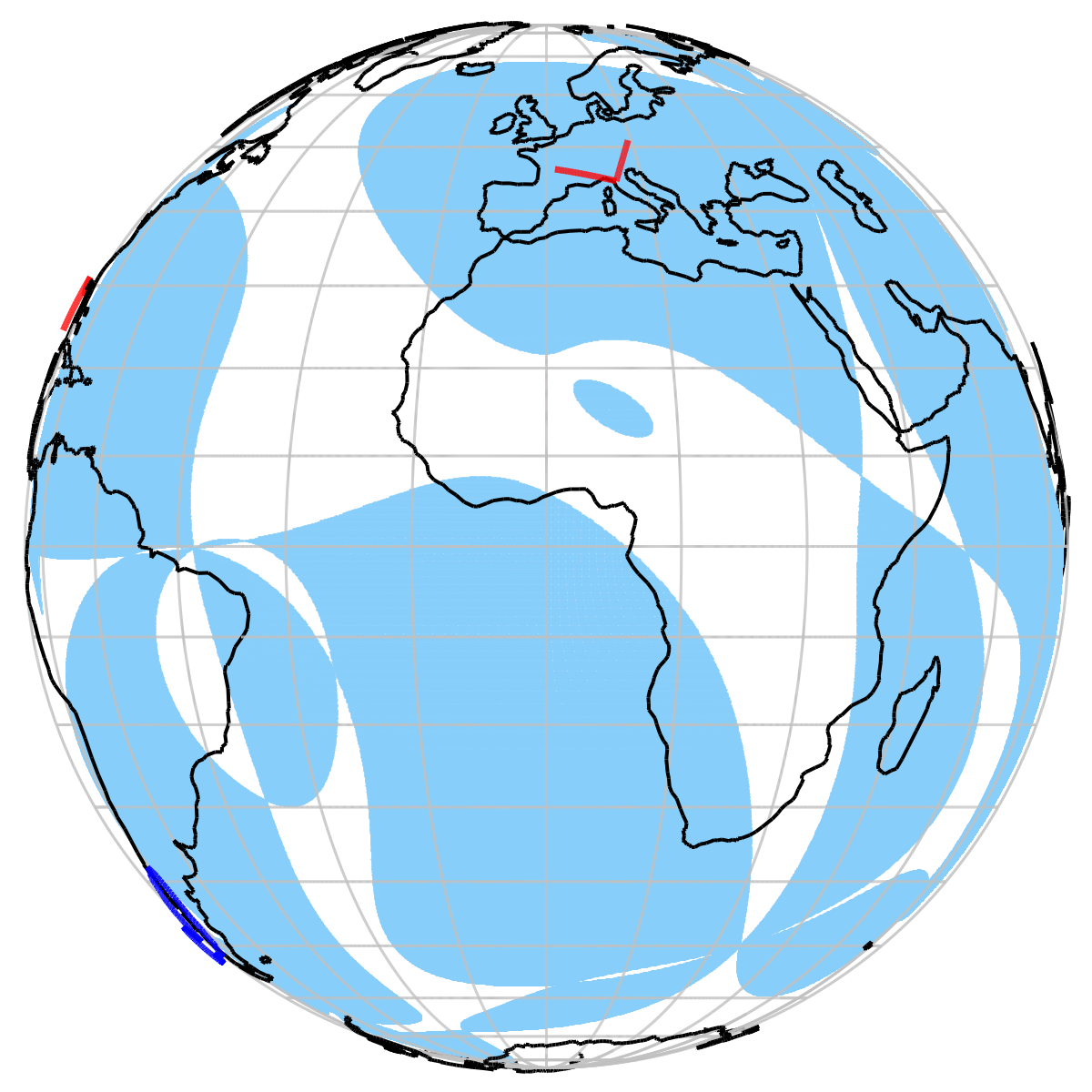} \\
    \includegraphics[width=0.4\textwidth]{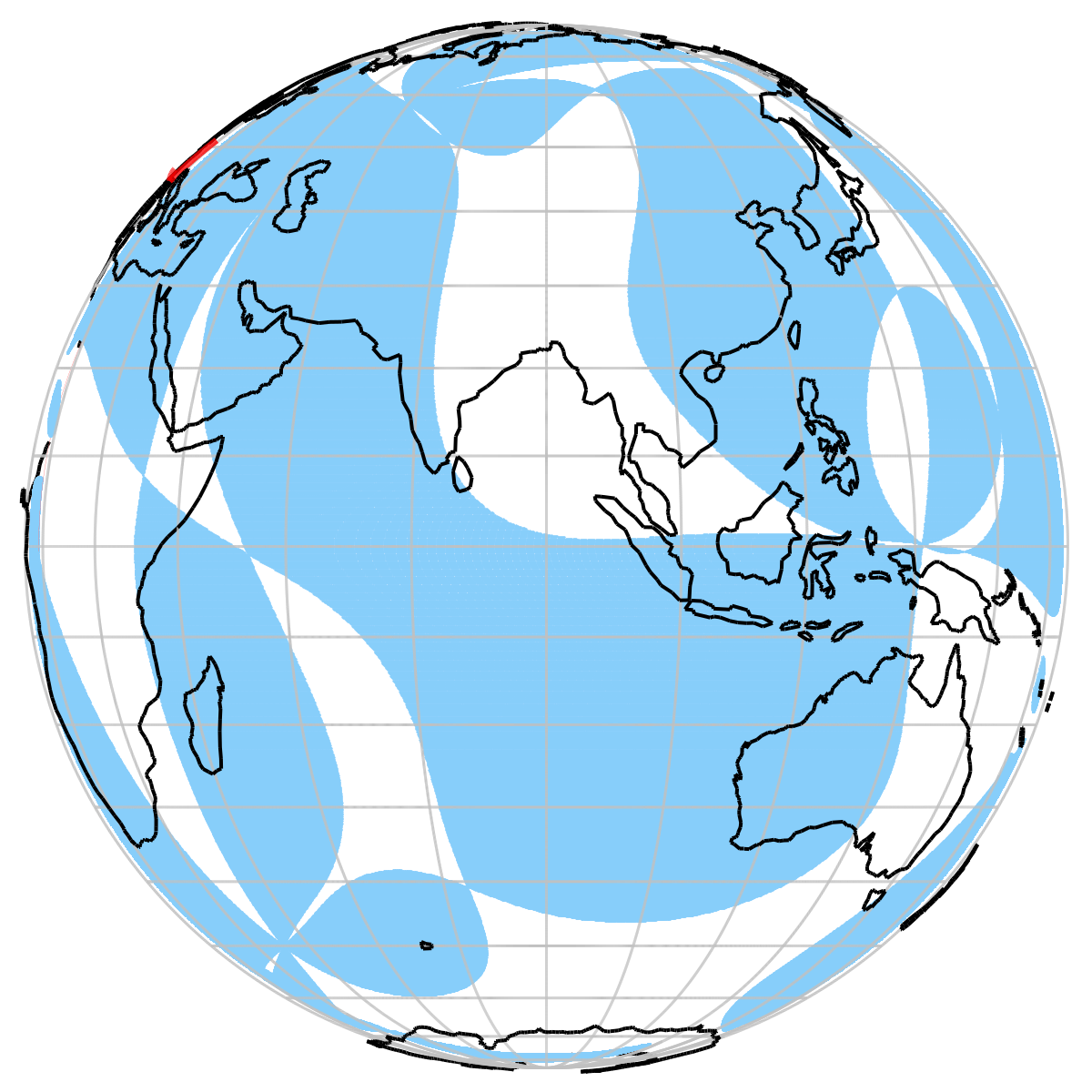} & \includegraphics[width=0.4\textwidth]{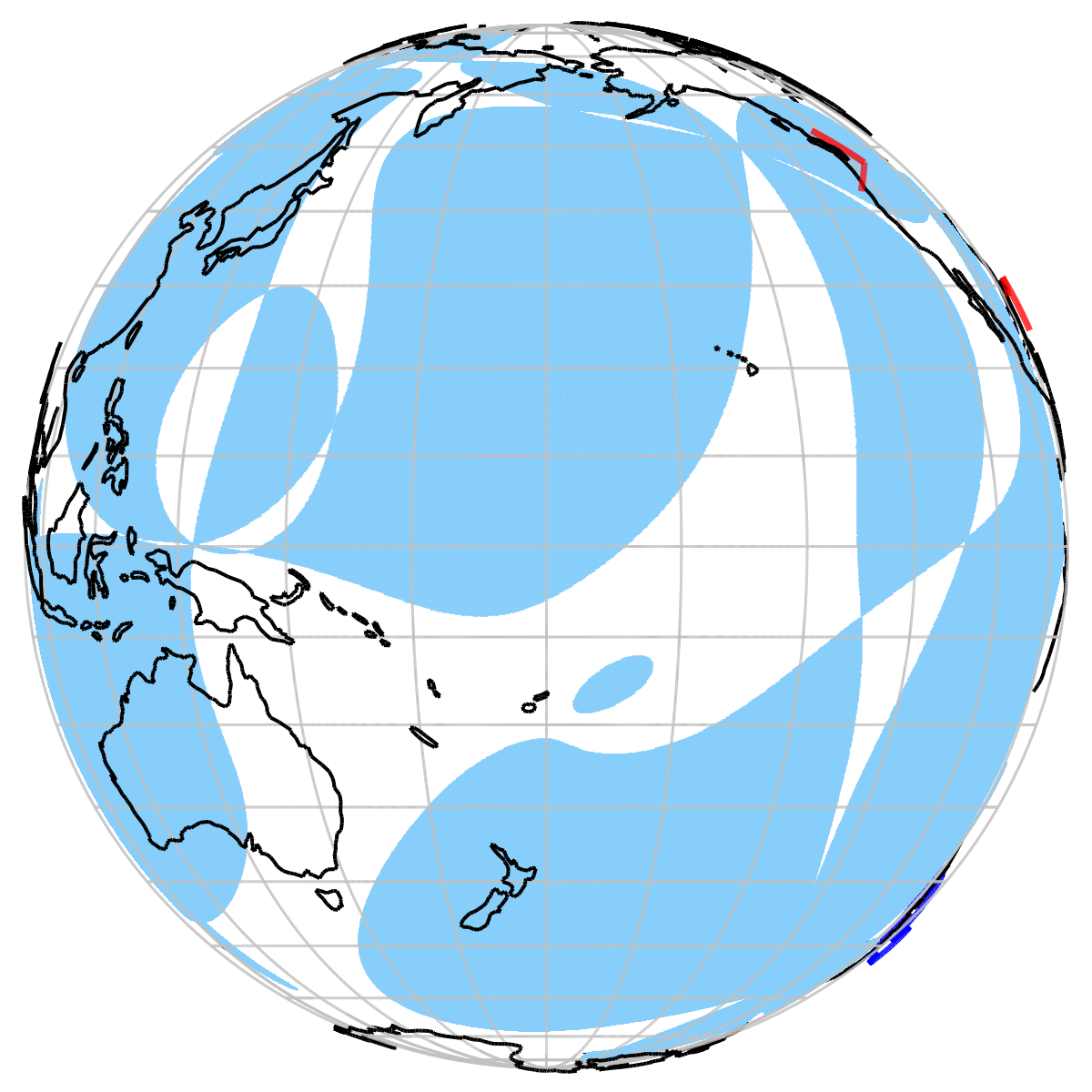}
  \end{tabular}
  \end{center}
 \caption{\label{f:skymaps} These plots show directions on the sky,
   relative to the currently-operating Earth-based gravitational wave
   detectors. The LIGO Hanford, LIGO Livingston, and Virgo detector
   arms are shown in red (not to scale).  Sky directions in which the
   triangle inequality is \emph{not} satisfied are shown in light
   blue; these are invariant under antipodal reflection.  The
   remaining sky directions (white) are those in which the pure
   vector model can mimic GR.  The dark blue path outlines the most
   likely region (at 90\% Bayesian confidence) from which the GW170814
   event originated.  In part of that region the pure vector model
   presented here can give detector signals in the three instruments
   which are the same as those from GR.}
\end{figure*}

At the moment, there are three sensitive interferometric gravitational
wave detectors in operation: LIGO Hanford, LIGO Livingston, and Virgo.
In Fig.~\ref{f:skymaps} we show the regions of the sky where this
triangle inequality is satisfied.  Pure vector theory sources in
those directions could mimic the spin-2 transverse traceless
perturbations of GR.  In particular, the area of the sky in which
GW170814 was likely located, overlaps these directions.

\section{Conclusion}
\label{s:conclusion}

What has been presented here is a cautionary note which demonstrates
that ruling out a pure vector polarization model, in the absence
of a theory which defines its behavior, is difficult.  We have
constructed a pure vector model which, for certain source
locations, can exactly mimic the signal predicted by GR in the LIGO
and Virgo detectors.  Of course the recent LIGO \cite{Abbott:2016blz,
  Abbott:2016nmj, Abbott:2017vtc} and LIGO-Virgo \cite{Abbott:2017oio,
  TheLIGOScientific:2017qsa} observations and many other experimental
tests indicate that GR is an excellent description of nature, so the
author disavows any belief in the contrived pure vector model
presented here.  It is only intended as a counterexample.

The pure vector model presented here is able to mimic
GR for a three-detector network because the polarization
mixture “rotates” as the gravitational wave propagates, so it
demonstrates that such polarization rotation is \emph{sufficient} to
mimic GR. While we have not proved that such rotation is
\emph{necessary} to mimic GR, that may well be true.

Counting free parameters suggests that it should be possible to
construct similar contrived pure vector models which would match
the prediction of GR for short-duration signals for networks
containing more than three detectors.  However the set of allowed sky
directions where this is possible would be further constrained (smaller total sky area).

At each detector location (labeled by $i$) this pure vector model
introduces an additional angle $\Delta \psi_i$.  So one might well
ask: does it have any predictive ability at all?  Can it be falsified?
The answer to both questions is yes.  Recall that the helicity of a
field is defined by the properties of the complex phase under
rotations \cite{Feynman:1996kb}. Imagine that we have a set of
co-located gravitational wave detectors, whose arms are all in the
same plane tangent to the Earth, but which are rotated with respect to
one another.  Then the helicity can be directly measured by comparing
the phase of the strain signals between the different instruments.
For example suppose that we had 360 co-located interferometric
detectors, each of which had its arms offset by $1^\circ$ from the
previous one, so that they spanned the entire $360^\circ$ circle in
$1^\circ$ increments.  In GR, the phase of the signal will go through
two complete cycles (at any instant in time) as you move around the
``circle of detectors''. In the pure vector theory there would be
only one cycle of phase change.  This thought-experiment demonstrates
that the pure vector model presented here does make falsifiable
predictions.

Similar reasoning indicates that this pure vector model would have
difficulty in mimicking GR for long-duration signals. This is because
the Earth's rotation changes the orientation of the source relative to
the detectors.  For example a detector located at the North Pole with
its arms in the plane tangent to the Earth would undergo a complete
$360^\circ$ rotation each day.  If a gravitational wave signal is
observed over that entire time span, then models like the one
presented here could then be constrained or falsified
\cite{Abbott:2017tlp}.

The pure vector model presented here has one undetermined phase
$\Delta \psi_i$ for each detector. We have shown that for specific
values of these phases, the model can exactly reproduce the observed
instrumental strains expected from general relativity.  It follows
that for a real instrument, in which detector noise makes it
impossible to identify the signal exactly, some range of these values
would be compatible with data.  So there is a temptation to use
Bayesian methods to compare the odds that this model (with uniform
uninformed priors on $\Delta \psi_i \in [0, 2 \pi]$) describes Nature,
versus the predictions of general relativity.  But this would miss the
point: the model presented here is merely intended to demonstrate how
difficult it is to rule something out, in the absence of a specific
predictive model.  One could compute odds ratios for the model
presented here, but the number of alternative models is only limited
by the imagination.  Without definite predictions, their odds ratios
are not computable.

\begin{acknowledgments}
The author is grateful for helpful discussions with Patrick Sutton and
Alan Weinstein, useful comments from B.S. Sathyaprakash and Andrea
Vicere, and assistance with the plots from Reed Essick, Oliver Bock,
and Heinz-Bernd Eggenstein.
\end{acknowledgments}

\section*{Appendix: Detector response tensors for LIGO and Virgo}

Here we list the quantities needed to easily reproduce
Fig.~\ref{f:skymaps}. We work in a right-handed Cartesian coordinate
system $[u,v,w]$ which is fixed to the Earth (so rotates with it) and
whose origin is at the center of the Earth.  The positive $u$-axis
passes through the line of Greenwich at zero longitude where it
crosses the equator, and the positive $w$-axis passes through the
North Pole. The Earth is assumed to be spherical.

The first quantities of interest are unit vectors $n$ pointing from
the center of the Earth to the detectors, which may be obtained from
the geographical information cataloged in \cite{GR-QC/9607075}. These
are
\begin{equation}
  \begin{aligned}
    n_{\rm LLO} & =    -0.01157 {\, \hat u} -0.86102 {\, \hat v} + 0.50844 {\, \hat w} \\
    n_{\rm LHO} & =    -0.33833 {\, \hat u} -0.60020 {\, \hat v} + 0.72477 {\, \hat w} \\
    n_{\rm Virgo} & = \,  \, \, \, \, 0.71169 {\, \hat u} + 0.13190 {\, \hat v} + 0.69000 {\, \hat w}.
  \end{aligned}
  \nonumber
\end{equation}
The next useful quantities are unit vectors $L_1$ and $L_2$ along the
two detector arms, derived from information in \cite{GR-QC/9607075}:
\begin{equation}
  \begin{aligned}
    \rm LLO:\;\;\;   & L_1 =  -0.95308 {\, \hat u} -0.14432 {\, \hat v}  -0.26609  {\, \hat w} \\
               & L_2 =  \,  \, \, \, \, 0.30249 {\, \hat u}  -0.48767 {\, \hat v}  -0.81895 {\, \hat w} \\
    \rm LHO: \;\;\;  & L_1 =  -0.23684  {\, \hat u} +  0.79971  {\, \hat v} +  0.55169 {\, \hat w} \\
               & L_2  = -0.91073 {\, \hat u}   + 0.01500 {\, \hat v}  -  0.41272 {\, \hat w} \\
    \rm Virgo:\;\;\; & L_1 = -0.70121 {\, \hat u} + 0.19275 {\, \hat v} + 0.68641 {\, \hat w} \\
               & L_2 = -0.04246 {\, \hat u} - 0.97234 {\, \hat v} + 0.22967 {\, \hat w}.
  \end{aligned}
    \nonumber
\end{equation}
Working in the same $u,v,w$ basis, the components of the detector response tensors
(\ref{e:detresponsetensor}) are given by:
\begin{equation}
  \begin{aligned}
  d^{ab}_{\rm LLO} = &
  \left[
    \begin{array}{rrr}
       \;\;\,0.81686 &  0.28507 &  0.50134\\
       \;\;\,0.28507 & -0.21699 & -0.36097\\
       \;\;\,0.50134 & -0.36097 & -0.59988
    \end{array}
    \right]\\
  d^{ab}_{\rm LHO} = &
  \left[
    \begin{array}{rrr}
    -0.77334 & -0.17575  & -0.50654\\
    -0.17575 &  0.63931  &  0.44739\\
    -0.50654 &  0.44739  &  0.13403
    \end{array}
    \right]\\
  d^{ab}_{\rm Virgo} = &
  \left[
    \begin{array}{rrr}
       \;\;\,0.48989 & -0.17644 & -0.47156\\
      -0.17644 & -0.90830 & 0.35562\\
      -0.47156 & 0.35562 & 0.41841
    \end{array}
    \right].\\
  \end{aligned}
    \nonumber
\end{equation}
Please note that since the Earth is not spherical, these quantities
differ from the actual values by fractional amounts of order
$10^{-3}$.  For our purposes, this may be neglected.

For a given sky location, the pure tensor polarization tensors $e^{+,
  \times}_{ab}$ in Eq.~(\ref{e:spin2polbasisvectors}) and the
pure vector polarization tensors $e^{L,R}_{ab}$ in
Eq.~(\ref{e:LRbasis}) may be easily constructed from the
right-handed basis
\begin{equation}
  \begin{aligned}
    x = & - \sin L {\, \hat u} + \cos L {\, \hat v}  \\
    y = &  \sin \ell \cos L {\, \hat u}  +  \sin \ell \sin  L {\, \hat v } -  \cos \ell  {\, \hat w} \\
    z = & -\cos \ell \cos L {\, \hat u}  -\cos \ell \sin L {\, \hat v} -  \sin \ell  {\, \hat w},
  \end{aligned}
  \nonumber
\end{equation}
where (positive) $\ell \in [-90^\circ, 90^\circ]$ is lattitude North
of the equator, and (positive) $L \in [-180^\circ, 180^\circ]$ is
longitude East of Greenwich. (The published version of this paper
has $\sin \ell$ and $\cos \ell$ interchanged in the equations defining
$x$, $y$, and $z$. This is corrected above.)

\newpage
\bibliography{references,missing}

\begin{thebibliography}{33}%
\makeatletter
\providecommand \@ifxundefined [1]{%
 \@ifx{#1\undefined}
}%
\providecommand \@ifnum [1]{%
 \ifnum #1\expandafter \@firstoftwo
 \else \expandafter \@secondoftwo
 \fi
}%
\providecommand \@ifx [1]{%
 \ifx #1\expandafter \@firstoftwo
 \else \expandafter \@secondoftwo
 \fi
}%
\providecommand \natexlab [1]{#1}%
\providecommand \enquote  [1]{``#1''}%
\providecommand \bibnamefont  [1]{#1}%
\providecommand \bibfnamefont [1]{#1}%
\providecommand \citenamefont [1]{#1}%
\providecommand \href@noop [0]{\@secondoftwo}%
\providecommand \href [0]{\begingroup \@sanitize@url \@href}%
\providecommand \@href[1]{\@@startlink{#1}\@@href}%
\providecommand \@@href[1]{\endgroup#1\@@endlink}%
\providecommand \@sanitize@url [0]{\catcode `\\12\catcode `\$12\catcode
  `\&12\catcode `\#12\catcode `\^12\catcode `\_12\catcode `\%12\relax}%
\providecommand \@@startlink[1]{}%
\providecommand \@@endlink[0]{}%
\providecommand \url  [0]{\begingroup\@sanitize@url \@url }%
\providecommand \@url [1]{\endgroup\@href {#1}{\urlprefix }}%
\providecommand \urlprefix  [0]{URL }%
\providecommand \Eprint [0]{\href }%
\providecommand \doibase [0]{http://dx.doi.org/}%
\providecommand \selectlanguage [0]{\@gobble}%
\providecommand \bibinfo  [0]{\@secondoftwo}%
\providecommand \bibfield  [0]{\@secondoftwo}%
\providecommand \translation [1]{[#1]}%
\providecommand \BibitemOpen [0]{}%
\providecommand \bibitemStop [0]{}%
\providecommand \bibitemNoStop [0]{.\EOS\space}%
\providecommand \EOS [0]{\spacefactor3000\relax}%
\providecommand \BibitemShut  [1]{\csname bibitem#1\endcsname}%
\let\auto@bib@innerbib\@empty
\bibitem [{\citenamefont {Einstein}(1916)}]{Einstein:1916cc}%
  \BibitemOpen
  \bibfield  {author} {\bibinfo {author} {\bibfnamefont {A.}~\bibnamefont
  {Einstein}},\ }\href@noop {} {\bibfield  {journal} {\bibinfo  {journal}
  {Sitzungsber. Preuss. Akad. Wiss. Phys. Math. Kl}\ ,\ \bibinfo {pages} {688}}
  (\bibinfo {year} {1916})}\BibitemShut {NoStop}%
\bibitem [{\citenamefont {Einstein}(1918)}]{Einstein:1918btx}%
  \BibitemOpen
  \bibfield  {author} {\bibinfo {author} {\bibfnamefont {A.}~\bibnamefont
  {Einstein}},\ }\href@noop {} {\bibfield  {journal} {\bibinfo  {journal}
  {Sitzungsber. Preuss. Akad. Wiss. Phys. Math. Kl}\ ,\ \bibinfo {pages} {154}}
  (\bibinfo {year} {1918})}\BibitemShut {NoStop}%
\bibitem [{\citenamefont {Wald}(1984)}]{Wald:1984rg}%
  \BibitemOpen
  \bibfield  {author} {\bibinfo {author} {\bibfnamefont {R.~M.}\ \bibnamefont
  {Wald}},\ }\href {\doibase 10.7208/chicago/9780226870373.001.0001} {\emph
  {\bibinfo {title} {{General Relativity}}}}\ (\bibinfo  {publisher} {Chicago
  University Press},\ \bibinfo {address} {Chicago},\ \bibinfo {year}
  {1984})\BibitemShut {NoStop}%
\bibitem [{\citenamefont {Misner}\ \emph {et~al.}(1973)\citenamefont {Misner},
  \citenamefont {Thorne},\ and\ \citenamefont {Wheeler}}]{Misner:1974qy}%
  \BibitemOpen
  \bibfield  {author} {\bibinfo {author} {\bibfnamefont {C.~W.}\ \bibnamefont
  {Misner}}, \bibinfo {author} {\bibfnamefont {K.~S.}\ \bibnamefont {Thorne}},
  \ and\ \bibinfo {author} {\bibfnamefont {J.~A.}\ \bibnamefont {Wheeler}},\
  }\href@noop {} {\emph {\bibinfo {title} {{Gravitation}}}}\ (\bibinfo
  {publisher} {W. H. Freeman},\ \bibinfo {address} {San Francisco},\ \bibinfo
  {year} {1973})\BibitemShut {NoStop}%
\bibitem [{\citenamefont {Schutz}(1985)}]{Schutz:1985jx}%
  \BibitemOpen
  \bibfield  {author} {\bibinfo {author} {\bibfnamefont {B.~F.}\ \bibnamefont
  {Schutz}},\ }\href@noop {} {\emph {\bibinfo {title} {{A First Course in
  General Relativity}}}}\ (\bibinfo  {publisher} {Cambridge University Press},\
  \bibinfo {address} {Cambridge, England},\ \bibinfo {year} {1985})\BibitemShut
  {NoStop}%
\bibitem [{\citenamefont {Will}(2014)}]{Will:2014kxa}%
  \BibitemOpen
  \bibfield  {author} {\bibinfo {author} {\bibfnamefont {C.~M.}\ \bibnamefont
  {Will}},\ }\href {\doibase 10.12942/lrr-2014-4} {\bibfield  {journal}
  {\bibinfo  {journal} {Living Rev. Relativity}\ }\textbf {\bibinfo {volume}
  {17}},\ \bibinfo {pages} {4} (\bibinfo {year} {2014})}\BibitemShut {NoStop}%
\bibitem [{\citenamefont {Aasi}\ \emph {et~al.}(2015)\citenamefont {Aasi} \emph
  {et~al.}}]{TheLIGOScientific:2014jea}%
  \BibitemOpen
  \bibfield  {author} {\bibinfo {author} {\bibfnamefont {J.}~\bibnamefont
  {Aasi}} \emph {et~al.} (\bibinfo {collaboration} {LIGO Scientific
  Collaboration}),\ }\href {\doibase 10.1088/0264-9381/32/7/074001} {\bibfield
  {journal} {\bibinfo  {journal} {Classical Quantum Gravity}\ }\textbf
  {\bibinfo {volume} {32}},\ \bibinfo {pages} {074001} (\bibinfo {year}
  {2015})}\BibitemShut {NoStop}%
\bibitem [{\citenamefont {Acernese}\ \emph {et~al.}(2015)\citenamefont
  {Acernese} \emph {et~al.}}]{TheVirgo:2014hva}%
  \BibitemOpen
  \bibfield  {author} {\bibinfo {author} {\bibfnamefont {F.}~\bibnamefont
  {Acernese}} \emph {et~al.} (\bibinfo {collaboration} {Virgo Collaboration}),\
  }\href {\doibase 10.1088/0264-9381/32/2/024001} {\bibfield  {journal}
  {\bibinfo  {journal} {Classical Quantum Gravity}\ }\textbf {\bibinfo {volume}
  {32}},\ \bibinfo {pages} {024001} (\bibinfo {year} {2015})}\BibitemShut
  {NoStop}%
\bibitem [{Note1()}]{Note1}%
  \BibitemOpen
  \bibinfo {note} {We assume that these detectors can only observe a single
  polarization mode, or equivalently only a single linear combination of
  polarization modes.}\BibitemShut {Stop}%
\bibitem [{\citenamefont {Cadonati}(2004)}]{Cadonati:2004ms}%
  \BibitemOpen
  \bibfield  {author} {\bibinfo {author} {\bibfnamefont {L.}~\bibnamefont
  {Cadonati}},\ }\href {\doibase 10.1088/0264-9381/21/20/012} {\bibfield
  {journal} {\bibinfo  {journal} {Classical Quantum Gravity}\ }\textbf
  {\bibinfo {volume} {21}},\ \bibinfo {pages} {S1695} (\bibinfo {year}
  {2004})}\BibitemShut {NoStop}%
\bibitem [{\citenamefont {Wen}\ and\ \citenamefont
  {Schutz}(2005)}]{Wen:2005ui}%
  \BibitemOpen
  \bibfield  {author} {\bibinfo {author} {\bibfnamefont {L.}~\bibnamefont
  {Wen}}\ and\ \bibinfo {author} {\bibfnamefont {B.~F.}\ \bibnamefont
  {Schutz}},\ }\href {\doibase 10.1088/0264-9381/22/18/S46} {\bibfield
  {journal} {\bibinfo  {journal} {Classical Quantum Gravity}\ }\textbf
  {\bibinfo {volume} {22}},\ \bibinfo {pages} {S1321} (\bibinfo {year}
  {2005})}\BibitemShut {NoStop}%
\bibitem [{\citenamefont {Isi}\ \emph {et~al.}(2017)\citenamefont {Isi},
  \citenamefont {Pitkin},\ and\ \citenamefont {Weinstein}}]{Isi:2017equ}%
  \BibitemOpen
  \bibfield  {author} {\bibinfo {author} {\bibfnamefont {M.}~\bibnamefont
  {Isi}}, \bibinfo {author} {\bibfnamefont {M.}~\bibnamefont {Pitkin}}, \ and\
  \bibinfo {author} {\bibfnamefont {A.~J.}\ \bibnamefont {Weinstein}},\ }\href
  {\doibase 10.1103/PhysRevD.96.042001} {\bibfield  {journal} {\bibinfo
  {journal} {Phys. Rev.}\ }\textbf {\bibinfo {volume} {D 96}},\ \bibinfo
  {pages} {042001} (\bibinfo {year} {2017})}\BibitemShut {NoStop}%
\bibitem [{\citenamefont {Eardley}\ \emph
  {et~al.}(1973{\natexlab{a}})\citenamefont {Eardley}, \citenamefont {Lee},
  \citenamefont {Lightman}, \citenamefont {Wagoner},\ and\ \citenamefont
  {Will}}]{Eardley:1973br}%
  \BibitemOpen
  \bibfield  {author} {\bibinfo {author} {\bibfnamefont {D.~M.}\ \bibnamefont
  {Eardley}}, \bibinfo {author} {\bibfnamefont {D.~L.}\ \bibnamefont {Lee}},
  \bibinfo {author} {\bibfnamefont {A.~P.}\ \bibnamefont {Lightman}}, \bibinfo
  {author} {\bibfnamefont {R.~V.}\ \bibnamefont {Wagoner}}, \ and\ \bibinfo
  {author} {\bibfnamefont {C.~M.}\ \bibnamefont {Will}},\ }\href {\doibase
  10.1103/PhysRevLett.30.884} {\bibfield  {journal} {\bibinfo  {journal} {Phys.
  Rev. Lett.}\ }\textbf {\bibinfo {volume} {30}},\ \bibinfo {pages} {884}
  (\bibinfo {year} {1973}{\natexlab{a}})}\BibitemShut {NoStop}%
\bibitem [{\citenamefont {Eardley}\ \emph
  {et~al.}(1973{\natexlab{b}})\citenamefont {Eardley}, \citenamefont {Lee},\
  and\ \citenamefont {Lightman}}]{Eardley:1974nw}%
  \BibitemOpen
  \bibfield  {author} {\bibinfo {author} {\bibfnamefont {D.~M.}\ \bibnamefont
  {Eardley}}, \bibinfo {author} {\bibfnamefont {D.~L.}\ \bibnamefont {Lee}}, \
  and\ \bibinfo {author} {\bibfnamefont {A.~P.}\ \bibnamefont {Lightman}},\
  }\href {\doibase 10.1103/PhysRevD.8.3308} {\bibfield  {journal} {\bibinfo
  {journal} {Phys. Rev.}\ }\textbf {\bibinfo {volume} {D 8}},\ \bibinfo {pages}
  {3308} (\bibinfo {year} {1973}{\natexlab{b}})}\BibitemShut {NoStop}%
\bibitem [{\citenamefont {Isi}\ \emph {et~al.}(2015)\citenamefont {Isi},
  \citenamefont {Weinstein}, \citenamefont {Mead},\ and\ \citenamefont
  {Pitkin}}]{Isi:2015cva}%
  \BibitemOpen
  \bibfield  {author} {\bibinfo {author} {\bibfnamefont {M.}~\bibnamefont
  {Isi}}, \bibinfo {author} {\bibfnamefont {A.~J.}\ \bibnamefont {Weinstein}},
  \bibinfo {author} {\bibfnamefont {C.}~\bibnamefont {Mead}}, \ and\ \bibinfo
  {author} {\bibfnamefont {M.}~\bibnamefont {Pitkin}},\ }\href {\doibase
  10.1103/PhysRevD.91.082002} {\bibfield  {journal} {\bibinfo  {journal} {Phys.
  Rev.}\ }\textbf {\bibinfo {volume} {D 91}},\ \bibinfo {pages} {082002}
  (\bibinfo {year} {2015})}\BibitemShut {NoStop}%
\bibitem [{\citenamefont {Nishizawa}\ \emph {et~al.}(2009)\citenamefont
  {Nishizawa}, \citenamefont {Taruya}, \citenamefont {Hayama}, \citenamefont
  {Kawamura},\ and\ \citenamefont {Sakagami}}]{Nishizawa:2009bf}%
  \BibitemOpen
  \bibfield  {author} {\bibinfo {author} {\bibfnamefont {A.}~\bibnamefont
  {Nishizawa}}, \bibinfo {author} {\bibfnamefont {A.}~\bibnamefont {Taruya}},
  \bibinfo {author} {\bibfnamefont {K.}~\bibnamefont {Hayama}}, \bibinfo
  {author} {\bibfnamefont {S.}~\bibnamefont {Kawamura}}, \ and\ \bibinfo
  {author} {\bibfnamefont {M.-a.}\ \bibnamefont {Sakagami}},\ }\href {\doibase
  10.1103/PhysRevD.79.082002} {\bibfield  {journal} {\bibinfo  {journal} {Phys.
  Rev.}\ }\textbf {\bibinfo {volume} {D 79}},\ \bibinfo {pages} {082002}
  (\bibinfo {year} {2009})}\BibitemShut {NoStop}%
\bibitem [{\citenamefont {Abbott}\ \emph
  {et~al.}(2017{\natexlab{a}})\citenamefont {Abbott} \emph
  {et~al.}}]{Abbott:2017oio}%
  \BibitemOpen
  \bibfield  {author} {\bibinfo {author} {\bibfnamefont {B.~P.}\ \bibnamefont
  {Abbott}} \emph {et~al.} (\bibinfo {collaboration} {Virgo, LIGO Scientific
  Collaborations}),\ }\href {\doibase 10.1103/PhysRevLett.119.141101}
  {\bibfield  {journal} {\bibinfo  {journal} {Phys. Rev. Lett.}\ }\textbf
  {\bibinfo {volume} {119}},\ \bibinfo {pages} {141101} (\bibinfo {year}
  {2017}{\natexlab{a}})}\BibitemShut {NoStop}%
\bibitem [{\citenamefont {Jackson}(1998)}]{Jackson:1998nia}%
  \BibitemOpen
  \bibfield  {author} {\bibinfo {author} {\bibfnamefont {J.~D.}\ \bibnamefont
  {Jackson}},\ }\href@noop {} {\emph {\bibinfo {title} {{Classical
  Electrodynamics}}}}\ (\bibinfo  {publisher} {Wiley, New York},\ \bibinfo
  {year} {1998})\BibitemShut {NoStop}%
\bibitem [{\citenamefont {Fowles}(1968)}]{Fowles:1968zgx}%
  \BibitemOpen
  \bibfield  {author} {\bibinfo {author} {\bibfnamefont {G.~R.}\ \bibnamefont
  {Fowles}},\ }\href
  {http://www.fulviofrisone.com/attachments/article/404/Introduction%20to%20Modern%20Optics.pdf}
  {\emph {\bibinfo {title} {{Introduction to Modern Optics}}}}\ (\bibinfo
  {publisher} {Holt, Rinehart and Winston},\ \bibinfo {address} {New York},\
  \bibinfo {year} {1968})\BibitemShut {NoStop}%
\bibitem [{\citenamefont {{Rybicki}}\ and\ \citenamefont
  {{Lightman}}(1979)}]{1979rpa..book.....R}%
  \BibitemOpen
  \bibfield  {author} {\bibinfo {author} {\bibfnamefont {G.~B.}\ \bibnamefont
  {{Rybicki}}}\ and\ \bibinfo {author} {\bibfnamefont {A.~P.}\ \bibnamefont
  {{Lightman}}},\ }\href@noop {} {\emph {\bibinfo {title} {{Radiative Processes
  in Astrophysics}}}}\ (\bibinfo  {publisher} {Wiley-Interscience, New York},\
  \bibinfo {year} {1979})\ p.\ \bibinfo {pages} {393}\BibitemShut {NoStop}%
\bibitem [{\citenamefont {{Born}}\ and\ \citenamefont
  {{Wolf}}(1999)}]{1999prop.book.....B}%
  \BibitemOpen
  \bibfield  {author} {\bibinfo {author} {\bibfnamefont {M.}~\bibnamefont
  {{Born}}}\ and\ \bibinfo {author} {\bibfnamefont {E.}~\bibnamefont
  {{Wolf}}},\ }\href@noop {} {\emph {\bibinfo {title} {{Principles of
  Optics}}}}\ (\bibinfo  {publisher} {Cambridge University Press, Cambridge,
  England},\ \bibinfo {year} {1999})\BibitemShut {NoStop}%
\bibitem [{Note2()}]{Note2}%
  \BibitemOpen
  \bibinfo {note} {This retarded time is not referred to the source, but rather
  to the time at which the corresponding wavefront passes the plane defined by
  $z=0$.}\BibitemShut {Stop}%
\bibitem [{Note3()}]{Note3}%
  \BibitemOpen
  \bibinfo {note} {Since we are in the weak field limit, indices may be freely
  raised and lowered.}\BibitemShut {Stop}%
\bibitem [{\citenamefont {Feynman}()}]{Feynman:1996kb}%
  \BibitemOpen
  \bibfield  {author} {\bibinfo {author} {\bibfnamefont {R.~P.}\ \bibnamefont
  {Feynman}},\ }\href@noop {} {\emph {\bibinfo {title} {{Feynman Lectures on
  Gravitation}}}},\ edited by\ \bibinfo {editor} {\bibfnamefont {F.~B.}\
  \bibnamefont {Morinigo}}, \bibinfo {editor} {\bibfnamefont {W.~G.}\
  \bibnamefont {Wagner}}, \ and\ \bibinfo {editor} {\bibfnamefont
  {B.}~\bibnamefont {Hatfield}}\ (\bibinfo  {publisher} {Addison-Wesley,
  Reading, MA, 1995), p. 232 (The advanced book program})\BibitemShut {NoStop}%
\bibitem [{\citenamefont {Allen}()}]{GR-QC/9607075}%
  \BibitemOpen
  \bibfield  {author} {\bibinfo {author} {\bibfnamefont {B.}~\bibnamefont
  {Allen}},\ }\href@noop {} {\ }\Eprint
  {http://arxiv.org/abs/arXiv:gr-qc/9607075} {arXiv:gr-qc/9607075} \BibitemShut
  {NoStop}%
\bibitem [{Note4()}]{Note4}%
  \BibitemOpen
  \bibinfo {note} {This basis rotates in phase at frequency ``one cycle per
  rotation'' ($\phi $) as opposed to (say) frequency ``two cycles per
  rotation'' ($2\phi $) or ``no cycles per rotation'' ($0\phi $). Since the
  polarization basis transforms as a rank-(0,2) tensor, this implies that the
  vector polarizations are helicity $\pm 1$ components of a spin-2 field. The
  name ``vector'' perturbations is therefore somewhat misleading, since this
  implies a spin-1 object, which is a rank-(0,1) or (1,0) tensor. However it
  has historically been used in this context, so we perpetuate the
  misnomer.}\BibitemShut {Stop}%
\bibitem [{\citenamefont {{Mandel}}\ and\ \citenamefont
  {{Wolf}}(1995)}]{1995ocqo.book.....M}%
  \BibitemOpen
  \bibfield  {author} {\bibinfo {author} {\bibfnamefont {L.}~\bibnamefont
  {{Mandel}}}\ and\ \bibinfo {author} {\bibfnamefont {E.}~\bibnamefont
  {{Wolf}}},\ }\href@noop {} {\emph {\bibinfo {title} {{Optical Coherence and
  Quantum Optics}}}}\ (\bibinfo  {publisher} {Cambridge University Press,
  Cambridge, England},\ \bibinfo {year} {1995})\BibitemShut {NoStop}%
\bibitem [{Note5()}]{Note5}%
  \BibitemOpen
  \bibinfo {note} {We assume here that $F_3$ does not vanish, meaning that the
  GR strain in the third detector is nonvanishing.}\BibitemShut {Stop}%
\bibitem [{\citenamefont {Abbott}\ \emph
  {et~al.}(2016{\natexlab{a}})\citenamefont {Abbott} \emph
  {et~al.}}]{Abbott:2016blz}%
  \BibitemOpen
  \bibfield  {author} {\bibinfo {author} {\bibfnamefont {B.~P.}\ \bibnamefont
  {Abbott}} \emph {et~al.} (\bibinfo {collaboration} {Virgo, LIGO Scientific
  Collaborations}),\ }\href {\doibase 10.1103/PhysRevLett.116.061102}
  {\bibfield  {journal} {\bibinfo  {journal} {Phys. Rev. Lett.}\ }\textbf
  {\bibinfo {volume} {116}},\ \bibinfo {pages} {061102} (\bibinfo {year}
  {2016}{\natexlab{a}})}\BibitemShut {NoStop}%
\bibitem [{\citenamefont {Abbott}\ \emph
  {et~al.}(2016{\natexlab{b}})\citenamefont {Abbott} \emph
  {et~al.}}]{Abbott:2016nmj}%
  \BibitemOpen
  \bibfield  {author} {\bibinfo {author} {\bibfnamefont {B.~P.}\ \bibnamefont
  {Abbott}} \emph {et~al.} (\bibinfo {collaboration} {Virgo, LIGO Scientific
  Collaborations}),\ }\href {\doibase 10.1103/PhysRevLett.116.241103}
  {\bibfield  {journal} {\bibinfo  {journal} {Phys. Rev. Lett.}\ }\textbf
  {\bibinfo {volume} {116}},\ \bibinfo {pages} {241103} (\bibinfo {year}
  {2016}{\natexlab{b}})}\BibitemShut {NoStop}%
\bibitem [{\citenamefont {Abbott}\ \emph
  {et~al.}(2017{\natexlab{b}})\citenamefont {Abbott} \emph
  {et~al.}}]{Abbott:2017vtc}%
  \BibitemOpen
  \bibfield  {author} {\bibinfo {author} {\bibfnamefont {B.~P.}\ \bibnamefont
  {Abbott}} \emph {et~al.} (\bibinfo {collaboration} {Virgo, LIGO Scientific
  Collaborations}),\ }\href {\doibase 10.1103/PhysRevLett.118.221101}
  {\bibfield  {journal} {\bibinfo  {journal} {Phys. Rev. Lett.}\ }\textbf
  {\bibinfo {volume} {118}},\ \bibinfo {pages} {221101} (\bibinfo {year}
  {2017}{\natexlab{b}})}\BibitemShut {NoStop}%
\bibitem [{\citenamefont {Abbott}\ \emph
  {et~al.}(2017{\natexlab{c}})\citenamefont {Abbott} \emph
  {et~al.}}]{TheLIGOScientific:2017qsa}%
  \BibitemOpen
  \bibfield  {author} {\bibinfo {author} {\bibfnamefont {B.~P.}\ \bibnamefont
  {Abbott}} \emph {et~al.} (\bibinfo {collaboration} {Virgo, LIGO Scientific
  Collaborations}),\ }\href {\doibase 10.1103/PhysRevLett.119.161101}
  {\bibfield  {journal} {\bibinfo  {journal} {Phys. Rev. Lett.}\ }\textbf
  {\bibinfo {volume} {119}},\ \bibinfo {pages} {161101} (\bibinfo {year}
  {2017}{\natexlab{c}})}\BibitemShut {NoStop}%
\bibitem [{\citenamefont {Abbott}\ \emph {et~al.}(2018)\citenamefont {Abbott}
  \emph {et~al.}}]{Abbott:2017tlp}%
  \BibitemOpen
  \bibfield  {author} {\bibinfo {author} {\bibfnamefont {B.~P.}\ \bibnamefont
  {Abbott}} \emph {et~al.} (\bibinfo {collaboration} {Virgo, LIGO Scientific
  Collaborations}),\ }\href {\doibase 10.1103/PhysRevLett.120.031104}
  {\bibfield  {journal} {\bibinfo  {journal} {Phys. Rev. Lett.}\ }\textbf
  {\bibinfo {volume} {120}},\ \bibinfo {pages} {031104} (\bibinfo {year}
  {2018})}\BibitemShut {NoStop}%
\end{thebibliography}%

\end{document}